\documentclass[pre,twocolumn,aps,showpacs,floatfix]{revtex4}

\usepackage{graphicx}
\usepackage[latin1]{inputenc}
\usepackage{dcolumn}
\usepackage{amsmath}
\usepackage{amsfonts}
\usepackage{amssymb}
\usepackage{epsf}
\usepackage{epsfig}
\usepackage{bm}

\newcommand{\be}{\begin{equation}}
\newcommand{\ee}{\end{equation}}
\newcommand{\ba}{\begin{eqnarray}}
\newcommand{\ea}{\end{eqnarray}}

\usepackage{soul}
\usepackage{cancel}

\begin{document}
\title{Reaction-Subdiffusion and Reaction-Superdiffusion Equations for
Evanescent Particles Performing Continuous Time Random Walks}
\author{E. Abad$^{1}$, S. B. Yuste$^{1}$, and Katja
Lindenberg$^{2}$}
\affiliation{$^{(1)}$
Departamento de F\'{\i}sica, Universidad de Extremadura,
E-06071 Badajoz, Spain\\
$^{(2)}$Department of Chemistry and Biochemistry, and BioCircuits Institute,
University of California San Diego, 9500 Gilman Drive, La Jolla, CA
92093-0340, USA}

\date{\today}

\begin{abstract}
Starting from a continuous time random walk (CTRW) model of particles that may
evanesce as they walk, our goal is to arrive at macroscopic
integro-differential equations for the probability density for a particle to be
found at point $\mathbf{r}$ at time $t$ given that it started its walk from
$\mathbf{r}_0$ at time $t=0$. The passage from the CTRW to an
integro-differential equation is well understood when the particles are not
evanescent. Depending on the distribution of stepping times and distances,
one arrives at standard macroscopic equations that may be ``normal''
(diffusion) or ``anomalous'' (subdiffusion and/or superdiffusion). The
macroscopic description becomes considerably more complicated and not
particularly intuitive if the particles can die during their walk.  While such
equations have been derived for specific cases, e.g., for
location-independent exponential evanescence, we present a more general
derivation valid under less stringent constraints than those found in the current
literature.

\end{abstract}

\pacs{02.50.Ey,82.40.-g,82.33.-z,05.90.+m}

\maketitle

\section{Introduction}
\label{sec:intro}
Continuous time random walks (CTRW) offer a sweeping framework to describe the
dynamics of particles whose motion may be ``anomalous.'' That is, in addition to
providing a way to describe the motion of diffusive particles at a more
microscopic level than, say, the more macroscopic diffusion equation, CTRWs can
also be used to describe particles whose motion is subdiffusive or
superdiffusive. The connection between CTRWs and the associated ``more
macroscopic'' description, which is in general an integro-differential
equation rather than a (partial) differential equation as in the case of
diffusion, has been firmly established.  The more macroscopic description is
often attractive because all the machinery of integro-differential equations
and associated boundary conditions can be brought to bear in their solution.
It should be noted that the starting point for problems that involve
subdiffusive or superdiffusive particles is often simply the
integro-differential equation itself. However, one
must exercise caution in simply starting with such a macroscopic description. It
should also be noted that these integro-differential equations can arise from
microscopic models other than a CTRW, the latter not always being the
appropriate framework.  An example is the motion of particles in random
landscapes of
various sorts, for example, one in which each site is associated with a
potential well whose depth is chosen from a random distribution. The walk in a
landscape of potential wells of random depths may be
normal in the sense that the escape from each well follows a Kramers law, but
the irregularity of the
landscape may give rise to trapping events that may slow down the progress of
the particles to the point of subdiffusion.

The situation becomes much more complicated when the moving particles
also undergo reactions ~\cite{Yadav, Eule}. This applies to
situations
such as reversible or irreversible conversion to a different species (e.g., $A
\to B$~\cite{Sagues,Henry,Schmidt} or $A \rightleftharpoons B$~\cite{Schmidt}),
reactions giving rise to propagating fronts (say, $A+B\to
C$~\cite{YusteAcedoLin}, $A+B\to 2A$~\cite{Campos}), binary reactions (e.g.,
$A+A\to 0$), or even spontaneous evanescence ($A\to 0$). Focusing on the latter
case, the description of the evolution of such evanescent particles at the
macroscopic level of a subdiffusion or superdiffusion equation, and the
investigation of the proper way to include the evanescence or reaction in such
equations, is a matter of continuing study and is usually carried out in the
context of rather specific models~\cite{Sagues,Langlands}.

In this paper we pursue this goal somewhat more generically: we consider the
deduction of subdiffusion or superdiffusion equations when the moving
particles are intrinsically evanescent. The death rate of the particles may in
general depend on location, time, how long a particle has spent at a
particular location -- one can think of a large variety of death scenarios.
Our presentation is based on a CTRW
formulation generalized to include particle evanescence.  One step of
the discussion concerns the way in which
evanescence can be built into a CTRW model, a choice which is not unique.  The
second
step is then to go from a CTRW model to a
subdiffusion or superdiffusion equation as appropriate.

In Sec.~\ref{ctrw} we start with some basic CTRW quantities and
relations, which we then use in Sec.~\ref{integraleq} to construct an integral
equation for the probability density of finding a particle at a
certain position at time $t$ given that it began its motion at a
different position
at time $t=0$. This integral equation includes the possibility that the particle
evanesces along the way, and is the starting point for the further derivation of
subdiffusive and superdiffusive equations. In Sec.~\ref{fractionaleqs} we
explicitly construct our fractional equations for the case of subdiffusion,
superdiffusion, and a mixture of the two.  We conclude with a summary and some
thoughts for the future in Sec.~\ref{conclusions}.

\section{Starting with a CTRW}
\label{ctrw}
Our goal is to arrive at integro-differential equations for the
probability density $w(\mathbf{r},t|\mathbf{r}_0,0)$ for a particle that
started its journey at point
$\mathbf{r}_0$ at time $t=0$ to be at $\mathbf{r}$ at time $t$.
No matter how we arrive at such an equation, the description of the motion of
the particles as a CTRW requires the introduction of the probability density
$\Psi(\mathbf{r}-\mathbf{r'},t-t')$ that a random walker jumps in a single
step from $\mathbf{r'}$ to $\mathbf{r}$ after waiting a time interval $t-t'$ at
the position $\mathbf{r'}$.  The dependence on only the difference of the two
position vectors reflects an assumption of spatial homogeneity for the jumping
mechanism. Associated
with this probability density is the usual waiting time probability density,
\begin{equation}
\psi(t)=\int d\mathbf{r'} \Psi(\mathbf{r'},t),
\end{equation}
and also the probability density for a single step displacement,
\begin{equation}
 \chi(\mathbf{r})=\int_0^{\infty} dt'\, \Psi(\mathbf{r},t').
\end{equation}
The ``normal'' or ``anomalous'' (subdiffusive or superdiffusive) character of
the process depends on the forms of the probability densities, which we will
specify later.

Since we deal with evanescent particles, we also introduce the
probability $\Xi(\mathbf{r},\mathbf{r'};t,t')$ that the particle has not died
spontaneously during the stepping process described by
$\Psi(\mathbf{r}-\mathbf{r'},t-t')$.
%
Also, finding the particle at $\mathbf{r}$ at time $t$ does not mean that it
jumped onto that location exactly at that time. In fact, it might have jumped
there at an earlier time and then just waited there without dying, or it might
have jumped there at an earlier time, returned any number of times before $t$,
and then waited there.  As a starting point in our route to
an integral equation for $w(\mathbf{r},t|\mathbf{r}_0,0)$,
it is thus appropriate to introduce
$q_{n}(\mathbf{r},t|\mathbf{r}_0,0)$, the probability density that a particle be
at position $\mathbf{r}$ at time $t$ exactly after making its $n$th jump, given
that it started its walk at time $t=0$ at position $\mathbf{r}_0$ \cite{Langlands}. This
initial condition is described by the equation
\begin{equation}
 q_0(\mathbf{r},t|\mathbf{r}_0,0)=\delta(\mathbf{r}-\mathbf{r}_0)\,\delta(t).
\label{initcond}
\end{equation}
From this follows the definition
\begin{equation}
 q(\mathbf{r},t|\mathbf{r}_0,0)\equiv\sum_{n=0}^{\infty}q_n(\mathbf{r},t|\mathbf
{r}_0,0),
\end{equation}
which is the probability density that the particle steps onto position
$\mathbf{r}$ exactly at time $t$ (regardless of how many times the
particle has stepped anywhere, including on $\mathbf{r}$, before this time).

It is straightforward to write integral equations for the probability densities
$q$ and $q_n$. For $q_n$ no further discussion beyond the definitions is
necessary to write the integral recurrence equation
\begin{eqnarray}
q_{n+1}(\mathbf{r},t|\mathbf{r}_0,0)&=&\int d\mathbf{r'}\int_0^t
dt' \,\Psi(\mathbf{r}-\mathbf{r'}
t-t')\nonumber\\
&\times& q_n(\mathbf{r'},t'|\mathbf{r}_0,0)\,\,\Xi(\mathbf{r},\mathbf{r'}
;t , t').
\label{mastereq}
\end{eqnarray}
In words, this says that the probability that the particle arrives at location
$\mathbf{r}$ at time $t$ on the $(n+1)$st step is equal to the probability that
it arrives at $\mathbf{r'}$ at an earlier time $t'$ on the $n$th step and then
steps from $\mathbf{r'}$ to $\mathbf{r}$ at time $t$, provided it does not die
in
the interval between these two steps. The usual integral equation without
evanescence is immediately recovered upon setting $\Xi$ equal to unity.
Summing this equation over step number $n$,
the integral equation
\begin{eqnarray}
q(\mathbf{r},t|\mathbf{r}_0,0)&=&\int d\mathbf{r'}\int_0^{t}
dt'\,\Psi(\mathbf{r}-\mathbf{r'},
t-t')\nonumber\\
&\times& q(\mathbf{r'},t'|\mathbf{r}_0,0)\,\Xi(\mathbf{r},\mathbf{r'};t,
t')
+q_0(\mathbf{r},t|\mathbf{r}_0,0)
\nonumber\\
\label{mastereq2}
\end{eqnarray}
immediately follows.

The more complex question is now how to go from these integral equations
to one for the desired probability density $w(\mathbf{r},t|\mathbf{r}_0,0)$ and
thence to a fractional diffusion equation. This is our pursuit in the next
section.

\section{Integral equation with evanescence}
\label{integraleq}

Before obtaining an integral equation from the CTRW setup of the last section,
we note that one could choose to write such an equation directly, for example,
\begin{eqnarray}
w(\mathbf{r},t|\mathbf{r}_0,0)&=&\int d\mathbf{r'}\,
\int_0^{t} dt'\,\ \varUpsilon(\mathbf{r},\mathbf{r'};t,t')\nonumber\\
\nonumber\\
&&\times w(\mathbf{r'},t'|\mathbf{r}_0,0)\,\varXi(\mathbf{r},\mathbf{r'};t,t')
\nonumber\\\nonumber\\
&+&\Phi(t)\,\varXi(\mathbf{r},\mathbf{r};t,0)\delta(\mathbf{r}-\mathbf{r}_0),
\label{reneq2}
\nonumber\\
\end{eqnarray}
where we have introduced
\begin{equation}
 \Phi(t)=1-\int_0^{t} dt'\,\psi(t')=\int_t^{\infty} dt'\,\psi(t'),
\end{equation}
the probability that the particle does not take a step in the entire time
interval up to time $t$. Note that we have made a point of using $\varXi$
rather than $\Xi$ for the function indicating that no death occurs, because in
general there is no reason to expect these two to be the same.  The function
$\Xi$ that appears in the integral equations obtained above is the probability
that the particle
does not die in a time interval exactly delineated by two steps (and none in
between), one taking it to location $\mathbf{r'}$ and the other to location
$\mathbf{r}$. On the other hand, $\varXi$ is the probability that the particle
does not die in a time interval $t-t'$ as it moves from $\mathbf{r'}$ to
$\mathbf{r}$, with no reference to steps. There is no a priori reason for these
two probabilities to be equal. If particles die at a constant rate
independent of position and of when steps take place, then these two
probabilities would be equal.  However, for example if particles die only
when they take a step, or, conversely, if particles are more likely to die
if they remain at one location, then these probabilities would not be
equal. Also, we have denoted the kernel of the integral
equation by the new symbol $\varUpsilon$ because a priori we do not know its
connection to $\Psi$. One might attempt reasonable guesses, but that is all so
far and in fact somewhat risky.

Our goal is to obtain an equation such as Eq.~(\ref{reneq2}) starting from a CTRW (rather
than just writing it down).
We therefore begin with Eq.~(\ref{mastereq2}) together with the exact relation
\begin{equation}
\label{relwq}
w(\mathbf{r},t|\mathbf{r}_0,0)=\int_0^t
dt'\,\Phi(t-t')q(\mathbf{r},t'|\mathbf{r}_0,0)\,\varXi(\mathbf{r},\mathbf{r};t,
t').
\end{equation}
This exact relation simply says that the probability density for the particle
to be at $\mathbf{r}$ at time $t$ is the probability that it stepped onto that
location at time $t' < t$ and then neither moved nor died until time $t$.
Earlier arrivals and returns are implicitly included in this relation.

In order to proceed toward an integral equation of the form (\ref{reneq2}) with
known kernel we find ourselves having to make two admittedly
debatable assumptions. One is that the
functions $\Xi$ and $\varXi$ are equal.
The other is that
this function can be written as a ratio of the form
\begin{equation}
 \Xi(\mathbf{r},\mathbf{r'};t,t')=\frac{\varphi(\mathbf{r},t)}{\varphi(\mathbf{
r'},t')}.
\label{speccase}
\end{equation}
The special case  of location-independent exponential
decay $\varphi(\mathbf{r},t) = \exp(-kt)$ (that is, the case of evanescence at
a constant rate, mentioned earlier), is that of Sokolov et
al.~\cite{Sagues}, and in this case
$\Xi(\mathbf{r},\mathbf{r'};t,t')=\exp[-k(t-t')]$.  This is the only choice for
which $\Xi$ depends on time only through the difference of its time arguments.
While Eq.~(\ref{speccase}) is limiting, it is more general than the cases that
have been treated in the literature.

To make use of the special form (\ref{speccase}) we start by making the
replacements $t\to t'$ and $t' \to t''$ in Eq.~(\ref{mastereq2}). We then
multiply by $\Phi(t-t')\,\Xi(\mathbf{r},\mathbf{r};t,t')$, integrate over $t'$,
and use relation (\ref{relwq}) to write
\begin{widetext}
\begin{eqnarray}
w(\mathbf{r},t|\mathbf{r}_0,0)&=&\int_0^t dt'\,\Phi(t-t')
\,\Xi(\mathbf{r},\mathbf{r};t,t')\left(\int d\mathbf{r'}\,\int_0^{t'}
dt''
\Psi(\mathbf{r}-\mathbf{r'},t'-t'')
q(\mathbf{r'},t''|\mathbf{r}_0,0)\,\Xi(\mathbf{r},\mathbf{r'};t',t'')\right)
\nonumber\\
&+&\int_0^t
dt'\,\Phi(t-t')\,\Xi(\mathbf{r},\mathbf{r};t,t')\,q_0(\mathbf{r},t'|\mathbf{r}_0
,0).
\label{wqeq}
\end{eqnarray}
\end{widetext}
This equation can be manipulated through a number of simple steps.  First, we
insert the special form Eq.~(\ref{speccase}) on the right side and divide both
sides of the equation by $\varphi(\mathbf{r},t)$. This leaves us with the
equation
\begin{widetext}
\begin{eqnarray}
\frac{w(\mathbf{r},t|\mathbf{r}_0,0)}
{\varphi(\mathbf{r},t)}&=&\int d\mathbf{r'}\,\int_0^t dt'\,\Phi(t-t')
\left(\int_0^{t'} dt''\,\Psi(\mathbf{r}-\mathbf{r'},t'-t'')
\frac{q(\mathbf{r'},t''|\mathbf{r}_0,0)}{\varphi(\mathbf{r'},t'')}\right)
\nonumber \\
&+&
\int_0^t
dt'\,\Phi(t-t')\frac{q_0(\mathbf{r},t'|\mathbf{r}_0,0)}{\varphi(\mathbf{r},t')}.
\label{inteq}
\end{eqnarray}
\end{widetext}
Next, we recognize that the first term on the right is a double convolution
with respect to time.  The memory kernels can then be exchanged, as can be
verified by taking Laplace transforms. Multiplying the rearranged equation by
$\varphi(\mathbf{r},t)$ it is then easy to see that
\begin{widetext}
\begin{eqnarray}
w(\mathbf{r},t|\mathbf{r}_0,0)
&=&\int d\mathbf{r'}\,\int_0^t dt'\,\Psi(\mathbf{r}-\mathbf{r'},t-t')
\frac{\varphi(\mathbf{r},t)}{\varphi(\mathbf{r'},t')}\left(\int_0^{t'}
dt''\,\Phi(t'-t'')
q(\mathbf{r'},t''|\mathbf{r}_0,0)\frac{\varphi(\mathbf{r'},t')}
{\varphi(\mathbf{r'},t'')}\right) \nonumber \\
&+&\int_0^t
dt'\,\Phi(t-t')q_0(\mathbf{r},t'|\mathbf{r}_0,0)\frac{\varphi(\mathbf{r},t)}
{\varphi(\mathbf{r},t')}.
\end{eqnarray}
\end{widetext}
Finally, using Eq.~(\ref{relwq}) and the initial condition (\ref{initcond}),
and again recalling the special form (\ref{speccase}) we arrive at the desired
integral equation
\begin{widetext}
\begin{equation}
w(\mathbf{r},t|\mathbf{r}_0,0)=\int d\mathbf{r'}\,
\int_0^{t} dt'\,\Psi(\mathbf{r}-\mathbf{r'},t-t')
w(\mathbf{r'},t'|\mathbf{r}_0,0)
\Xi(\mathbf{r},\mathbf{r'}; t,t')
+\Phi(t)\Xi(\mathbf{r},\mathbf{r};t,0)\delta(\mathbf{r}
-\mathbf{r}_0).
\label{theintegralequation}
\end{equation}
\end{widetext}

Equation~(\ref{theintegralequation}) is the starting point for the derivation
of various fractional diffusion equations for different forms of the single
step probability density $\Psi(\mathbf{r}-\mathbf{r'},t-t')$ of the underlying
CTRW. It is therefore a centerpiece of this work. Note that the special form
(\ref{speccase}) has led to a kernel in the
integral equation that is precisely this single step probability density even
though the times $t$ and $t'$ are not necessarily associated with jumping
times. We also stress once again that in addition to the simplification
(\ref{speccase}) we have assumed the equality of the survival functions $\Xi$
and $\varXi$. Equation~(\ref{theintegralequation}) has a clear physical
interpretation: it considers all possible ways for a particle to be at point
$\mathbf{r}$ at time $t$ by looking at the positions $\mathbf{r'}$ at prior
times $t'$ and then tracking
their subsequent arrival at the desired point. If the point under
consideration is the initial position,
the equation tracks the possibility that the particle has not moved by time
$t$. In the language of Hughes~\cite{Hughes1} (Sec. 3.2.8),
Eq.~(\ref{theintegralequation}) corresponds to a ``partition over
the last step.'' The equation counts only those particles that do not evanesce
in the process.

Finally, we shall implement one additional simplifying assumption widely adopted
in
the
literature, namely,that the waiting time and jump displacement distributions
are mutually independent, so we can write
\begin{equation}
\label{jumpdens}
\Psi(\mathbf{r},t)=\psi(t)\chi(\mathbf{r}).
\end{equation}
Different fractional diffusion equations then arise depending on the behaviors
of the tails of these distributions. We proceed to present various cases in the
next section. Our derivations closely follow known results presented
in a number of helpful review sources such as the reports of Metzler and
Klafter~\cite{MetzlerKlafterPhysReport,MetzlerKlafterRestaurant}
and a recent multiauthored
compendium on anomalous processes~\cite{AnotransBook}. Our main purpose here is
to add evanescence to the mix and to determine how the evanescence ``reaction''
enters these equations.  In particular, when $\psi(t)$ has long tails and
$\chi({\mathbf{r}})$ does not, we will arrive at a fractional subdiffusion
equation.  When, on the other hand, $\chi({\mathbf{r}})$ has long tails but
$\psi(t)$ does not, we arrive at a superdiffusive equation.  The most
``anomalous'' case occurs when both have long tails, which leads to a
bifractional equation.

We end this section with a practical consideration.
Instead of working with the probability density of interest,
$w(\mathbf{r},t|\mathbf{r}_0,0)$, it turns out to be more convenient to work
with a ratio introduced earlier,
\begin{equation}
\label{vartrafo}
\eta(\mathbf{r},t|\mathbf{r}_0,0)=
\frac{w(\mathbf{r},t|\mathbf{r}_0,0)}{\varphi(\mathbf{r},t)}.
\end{equation}
Dividing Eq.~(\ref{theintegralequation}) by $\varphi(\mathbf{r},t)$ and taking
the Fourier (for space)-Laplace (for time) transform, we find
\begin{equation}
\hat{\tilde{\eta}}(\mathbf{q},u)=\hat{\chi}(\mathbf{q})\tilde{\psi}(u)
\hat{\tilde{\eta}}(\mathbf{q},u)+\tilde{\Phi}(u)\,
\frac{e^{i\mathbf{q}\cdot\mathbf{r}_0}}{\varphi(\mathbf{r}_0,0)}.
\label{howcomenolabel}
\end{equation}
Together with the relation
\begin{equation}
\tilde{\Phi}(u)=\frac{1}{u}-\frac{\tilde{\psi}(u)}{u}
\end{equation}
and the convolution theorems for both Fourier and Laplace transforms, we
arrive at the Fourier-Laplace transformed reaction-diffusion equation with
evanescence, equivalent to Eq.~(\ref{theintegralequation}) when the memory
kernel can be factorized as in Eq.~(\ref{jumpdens}),
\begin{widetext}
\begin{equation}
\label{fourlapeq}
u\hat{\tilde{\eta}}(\mathbf{q},u)=u
\hat{\chi}(\mathbf{q})\tilde{\psi}(u)\hat{\tilde{\eta}}(\mathbf{q},u)
+\left(1-\tilde{\psi}(u)\right)
\frac{e^{i\mathbf{q}\cdot\mathbf{r}_0}}{\varphi(\mathbf{r}_0,0)}.
\end{equation}
\end{widetext}

\section{Fractional diffusion equations with evanescence}
\label{fractionaleqs}
In this section we proceed to deduce the fractional diffusion equations
appropriate for long-tailed waiting time distributions, for long-tailed jump
distance distributions, and for both simultaneously.

\subsection{Fractional subdiffusive equation with evanescence}
\label{sec:ecuSubdifu}

Subdiffusion is characterized by a waiting time distribution with a tail
so long that it lacks integer moments, that is,
\begin{equation}
\psi(t)\sim  \frac{\kappa}{\tau_D}\left(\frac{t}{\tau_D}\right)^{-\gamma-1},
\label{wait}
\end{equation}
with $0<\gamma<1$. Here $\kappa$ is a dimensionless constant and $\tau_D$ is a
characteristic mesoscopic time (but not a first moment).
The small-$u$
behavior of the Laplace transform of the waiting time distribution reads
\begin{equation}
\label{laptranspsi}
\tilde{\psi}(u)\sim 1-{\cal A} u^\gamma ,
\end{equation}
with ${\cal A}=\gamma^{-1}\kappa\Gamma(1-\gamma)\tau_D^\gamma$.

We take the jump distance distribution to be ``normal,'' that is, it has finite
moments. In this case,
we are interested in small values of $q=|\mathbf{q}|$, for which one can
expand the Fourier transform of the jump displacement distribution and retain
only the first two terms,
\begin{equation}
\label{fourtransdens}
\hat{\chi}(\mathbf{q})\sim 1-\frac{\sigma^2\mathbf{q}^2}{2}+{\cal
O}(\mathbf{q}^4),
\end{equation}
where the second moment
\begin{equation}
\sigma^2=\int d\mathbf{r}\, \mathbf{r}^2 \chi(\mathbf{r})
\end{equation}
is assumed to be finite.

Substitution of the expansions~(\ref{laptranspsi}) and (\ref{fourtransdens})
into Eq.~(\ref{fourlapeq}) and neglect of a term of ${\cal O}(u\mathbf{q}^2)$
(which is unimportant in the asymptotic regime of small wave vectors and low
frequencies) leaves us with
\begin{equation}
\label{fullfourlap2}
u\hat{\tilde{\eta}}(\mathbf{q},u)-
\frac{e^{i\mathbf{q}\cdot\mathbf{r}_0}}{\varphi(\mathbf{r}_0,0)}
=-u^{1-\gamma}\frac{\sigma^2\mathbf{q}^2}{2{\cal
A}}\hat{\tilde{\eta}}(\mathbf{q},u).
\end{equation}
Laplace and Fourier inversion then yield
\begin{equation}
\label{purefrde}
\frac{\partial \eta(\mathbf{r},t|\mathbf{r}_0,0)}{\partial t}=
K_\gamma ~_{0}\,{\cal D}_t^{1-\gamma}\nabla^2_\mathbf{r}\,
\eta(\mathbf{r},t|\mathbf{r}_0,0),
\end{equation}
where we have introduced the anomalous diffusion coefficient
\be
K_\gamma=\frac{\sigma^2}{2{\cal A}}.
\ee
The integrodifferential operator ${\cal D}_t^{1-\gamma}$ acting on $y(t)$ is
defined as the inverse Laplace transform of $u^{1-\gamma}\tilde{y}(u)$,
\be
{\cal L}_{u\to t}^{-1}\left\{u^{1-\gamma}\tilde{y}(u)\right\}=~_{0}\,{\cal
D}_t^{1-\gamma}
y(t),
\ee
and is closely related to the Riemann-Liouville operator $_{0}D_t^{1-\gamma}$
defined by
\be
~_{0}D_t^{1-\gamma} f(\mathbf{r},t)
=\frac{1}{\Gamma(\gamma)}\frac{\partial}{\partial t}
\int_0^t dt'\, \frac{f(\mathbf{r},t')}{(t-t')^{1-\gamma}}.
\ee
In fact, both operators ${\cal D}_t^{1-\gamma}$ and $_{0}D_t^{1-\gamma}$ are
the same when applied to sufficiently regular functions $f(t)$ as determined by
the condition $\lim_{t\to 0} \int_0^t d\tau (t-\tau)^{\gamma-1} f(\tau)
=0$ (see Pgs. 384 and 118 in ~\cite{AnotransBook}). This condition is satisfied
for all situations of interest here.

In terms of the probability density $w$, Eq.~(\ref{purefrde})
explicitly yields the fractional reaction-subdiffusion equation
\begin{widetext}
\be
\label{fracdifreeqXX}
\frac{\partial w(\mathbf{r},t|\mathbf{r}_0,0)}
{\partial t}=\varphi(\mathbf{r},t)\, K_{\gamma}
~_{0}\,{\cal D}_t^{1-\gamma}
\nabla_{\mathbf{r}}^2 \frac{1}{\varphi(\mathbf{r},t)}
w(\mathbf{r},t|\mathbf{r}_0,0)
+\frac{\dot{\varphi}(\mathbf{r},t)}{\varphi(\mathbf{r},t)}
w(\mathbf{r},t|\mathbf{r}_0,0).
\ee
\end{widetext}
Equation~(\ref{fracdifreeqXX}) is the first mesoscopic highlight of our
paper in that all further results for subdiffusion are obtained as special
cases. It is perhaps the most general fractional subdiffusion equation
associated with a CTRW with evanescence for a single-species system obtained to
date. It is not as general as one might hope
because of the rather stringent condition~(\ref{speccase}), but to the best of
our knowledge it does include
the equations currently in the literature.
The results of Sokolov et
al.~\cite{Sagues} and the explicit results in Henry et al.~\cite{Langlands} are
recovered if we set $\varphi(\mathbf{r},t) \propto \exp(-kt)$.

Somewhat more elaborate is the
connection with the recent work of Fedotov~\cite{FedotovCondMat}, which in turn
recovers the more general results in~\cite{Langlands}.  In the language of
Fedotov translated to our work, his choice corresponds to the particular
selection (in one dimension)
\be
\varphi(x,t)=\exp\left\{  \int_\tau^t r[\rho(x,t')] dt' \right\}.
\label{relation}
\ee
Here in Fedotov's language the chemical reaction responsible for the
evanescence is assumed to follow the law of mass action so that the reaction
term is of the form $r(\rho)\rho$, that is,
\be
\label{decayform}
\left[\frac{\dot \rho}{\rho}\right]_\text{Reaction}=r[\rho(x,t)].
\ee
In turn, $\rho(x,t)$ is the density
of (surviving) particles at point $x$ at time $t$, which is related to our
probability density via an integration over all initial positions,
\be
\rho(x,t)=\int dx_0 w(x,t;x_0,0).
\ee
The time $\tau$ in Eq.~(\ref{relation}) is an arbitrary reference time that can
be chosen to be zero.
Specifically, the choice $\tau=0$, substitution of Eq.~(\ref{relation})
into our general equation~(\ref{fracdifreeqXX}), and subsequent integration
over the initial condition leads exactly to Eq.~(19) in
Ref.~\onlinecite{FedotovCondMat}, which is thus again a special case of our
more general
formalism. The result is especially noteworthy because the probability of
spontaneous death is not set a priori but depends on the changing density
itself, This leads to an interesting complex nonlinear problem.

As a final note in this section we mention that under the same conditions that
led to our subdiffusive fractional equation~(\ref{fracdifreeqXX}) with a
Riemann-Liouville operator, we can arrive at an equivalent fractional
subdiffusion equation of Caputo form.  It turns out to be
\be
\frac{\partial^\gamma }{\partial t^\gamma}
\frac{w(\mathbf{r},t|\mathbf{r}_0,0)}{\varphi(\mathbf{r},t)}=
K_\gamma  \nabla^2_\mathbf{r}\,
\frac{w(\mathbf{r},t|\mathbf{r}_0,0)}{\varphi(\mathbf{r},t)}.
\ee
This is in general not our preferred choice because of the difficulties in
carrying out the Caputo fractional derivative of a product.

\subsection{Fractional superdiffusive equation with evanescence}
In the previous subsection we dealt with a waiting time distribution with long
tails together with a ``normal'' distribution of displacements, and
the end result was a fractional subdiffusion equation.  In this subsection we
consider a ``normal'' waiting time distribution along with a distribution of
displacements that has long tails.  This will lead to a superdiffusive equation.

We thus consider a waiting time distribution with a finite mean waiting time
$\tau$ between steps, so that its Laplace transform at small argument
(corresponding to long times) behaves as
\be
\label{laptranspsiNormal}
\tilde{\psi}(u)\sim 1-\tau u.
\ee
For the jump length distribution we assume an inverse power law behavior,
\be
\chi({\mathbf{r}})\sim \frac{\sigma^\mu}{r^{1+\mu}}
\label{jump}
\ee
with $1\le \mu\le 2$ and $r=|{\mathbf{r}}|$.  Its variance diverges, and its
Fourier transform is
\be
\label{FourTranschiAnoma}
\hat{\chi}({\mathbf{q}})\sim 1-\sigma^\mu q^\mu,
\ee
where $q=|{\mathbf{q}}|$.  The steps to follow are now straightforward. Again,
it turns out to be convenient to work with Eq.~(\ref{howcomenolabel}).
Substitution of the above expansions into this equation, retention
of leading terms, and some simple algebra leads to
\be
u \hat{\tilde{\eta}}({\mathbf{q}},u)-  \frac{e^{i {\mathbf{q}} \cdot
{\mathbf{r}}_0}}{\varphi({\mathbf{r}}_0,0)}=
 -  \frac{\sigma^\mu}{\tau} q^\mu \hat{\tilde{\eta}}({\mathbf{q}},u),
 \ee
or, inverting the time Laplace transfrom,
\be
\frac{\partial }{\partial t}\hat \eta({\mathbf{q}},t)  =
   -  \frac{\sigma^\mu}{\tau} q^\mu \hat{ \eta}({\mathbf{q}},t).
\ee

Let $\partial^\mu/\partial r^\mu$  be the operator defined by the following Fourier transform
property,
 \be
 \label{defiOperator}
 \mathfrak{F}\left\{\frac{\partial^\mu f(\mathbf{r})}{\partial r^\mu}\right\}
=-q^\mu f(\mathbf{q}).
 \ee
 We can thus write
\be
\label{purefrdeSuperdifu}
\frac{\partial \eta({\mathbf{r}},t|{\mathbf{r}}_0,0)}{\partial t}=
K  \frac{\partial^\mu  }{\partial r^\mu} \eta({\mathbf{r}},t|{\mathbf{r}}_0,0),
\ee
where we have introduced the anomalous diffusion coefficient
\be
K =\frac{\sigma^\mu}{\tau}
\ee
For a one-dimensional system $\partial^\mu/\partial r^\mu=\partial^\mu/\partial
x^\mu$ is the  Riesz operator
\cite{MetzlerKlafterPhysReport,MetzlerKlafterRestaurant,AnotransBook}.

Undoing the variable transformation (\ref{vartrafo}), we finally
arrive at the fractional reaction-superdiffusion equation,
\begin{widetext}
\be
\label{fracdifreeq1}
\frac{\partial w(\mathbf{r},t|\mathbf{r}_0,0)}
{\partial t}=\varphi(\mathbf{r},t)\, K
\frac{\partial^\mu  }{\partial r^\mu}
\frac{1}{\varphi(\mathbf{r},t)}
w(\mathbf{r},t|\mathbf{r}_0,0)
+\frac{\dot{\varphi}(\mathbf{r},t)}{\varphi(\mathbf{r},t)}
w(\mathbf{r},t|\mathbf{r}_0,0).
\ee
\end{widetext}
This is the second important mesoscopic result of our paper, namely, the
derivation of a reaction-superdiffusion fractional equation starting from a
CTRW.

 \subsection{Bifractional equation with evanescence}
 \label{sec:ecuBifraccional}
Finally, in this subsection we combine subdiffusion and superdiffusion in that
we choose a waiting time distribution that lacks finite moments (and thus leads
to subdiffusion by itself) with a jump distance distribution that also lacks
moments (and thus leads to superdiffusion by itself).  Our methodology directly
lends itself to this combination.

We choose the waiting time distribution of Eq.~(\ref{wait})
whose Laplace transform is given in Eq.~(\ref{laptranspsi}), and the jump
distribution
Eq.~(\ref{jump})
whose Fourier transform is given in Eq.~(\ref{FourTranschiAnoma}).
The steps to follow are now essentially the same as in the previous sections,
with appropriate care given to the retention of the leading contributions. After
some algebra we find
\be
\frac{\partial \eta({\mathbf{r}},t|{\mathbf{r}}_0,0)}{\partial t}=
K   ~_{0}{\cal D}_t^{1-\gamma} \frac{\partial^\mu  }{\partial r^\mu}
\eta({\mathbf{r}},t|{\mathbf{r}}_0,0).
\ee
Undoing the variable transformation (\ref{vartrafo}), we finally
obtain the fractional reaction-sub/super-diffusion equation,
\begin{widetext}
\be
\label{fracdifreeq2}
\frac{\partial w(\mathbf{r},t|\mathbf{r}_0,0)}
{\partial t}=\varphi(\mathbf{r},t)\, K
~_{0}{\cal D}_t^{1-\gamma}
\frac{\partial^\mu  }{\partial r^\mu}
\frac{1}{\varphi(\mathbf{r},t)}
w(\mathbf{r},t|\mathbf{r}_0,0)
+\frac{\dot{\varphi}(\mathbf{r},t)}{\varphi(\mathbf{r},t)}
w(\mathbf{r},t|\mathbf{r}_0,0).
\ee
\end{widetext}
This is our third mesoscopic result and is unique in that it combines both
subdiffusion and superdiffusion in a single equation.


\section{Summary and Outlook}
\label{conclusions}
In this paper we have approached the problem of describing the evolution
equation of particles that move in a medium in which they can
also die in a medium as they
move. The model is based on a CTRW description of the motion of the
particles. The motion may be
anomalous (subdiffusive or
superdiffusive) and the particles may die at a rate that can depend on
position as well as time. We are able to capture the models that have been
explicitly considered in the literature, e.g., the space-independent exponential
evanescence model of Refs.~\cite{Sagues} and \cite{Langlands}, but our model
can also capture complicated position
dependences of the evanescent behavior such as that of the model of
Fedotov~\cite{FedotovCondMat} that render the problem nonlinear. We
confirm in a more general way than had been established previously the
known result that the interplay of the (anomalous) motion and the evanescence is
quite complex and that in general it can not be represented as the sum of two
processes
the way it can in normal reaction-diffusion scenarios. Having said this, we
note as an aside that in certain
cases (e.g., that of evanescence at a constant rate $r=k$) it is
possible to
reduce the reaction-subdiffusion problem to a pure subdiffusion problem by a
proper transformation \cite{FedotovCondMat}, much in the spirit of
Danckwerts' solution for
the problem of classical diffusion with a linear reaction \cite{crank}.

We started by constructing an integral equation for the probability density of
finding a
(surviving) particle at a location $\mathbf{r}$ at time $t$ given that it
stepped on location $\mathbf{r}_0$ at time $t=0$. To proceed from this
CTRW-based equation to the fractional equations,
we found it necessary to make
some specific assumptions about the
form of the rate of evanescence. The kernel of the integral
equation under these conditions is simply related to the single-step jump
probability density of the underlying CTRW.  In spite of the constraints,
our models include as special cases all the explicit models that have been
presented in the literature. All those, case by
case~\cite{Sagues,Langlands,FedotovCondMat}, have been presented with very
specific physical contexts in mind.

Once we have arrived at an integral equation, the derivation of various
fractional
diffusion equations relies on fairly standard procedures dictated by the form
of the single-step probability properties, except that we have added
evanescence to the picture and are thus able to see the complex interplay of
motion and evanescence, at least under our assumptions.  Eventually we
hope to be able to relax some of our more stringent assumptions.  We also
hope to be able to include other particle loss mechanisms such as bimolecular reactions
in our scheme, possibly at the expense of introducing some kind of
mean-field assumption
to deal with the complexity arising from effects of cooperativity.
The existing formalism can be adapted
to some situations where instead of evanescence we have
particle sources, or perhaps sources and sinks simultaneously.
However, in some cases particle sources require special scrutiny.
For example,
special care is needed when dealing with particles that give rise to offspring
because one must specify the rules surrounding the location and time of
creation of new particles, especially when dealing with jump and waiting time
distributions that have long tails.
These are
all plans for future work. Our most immediate plans are to apply our results to
the problem of the survival probability of a target that is surrounded by a
$d$-dimensional sea of evanescent traps whose motion may be subdiffusive if
the waiting times for motion are too long, or superdiffusive if the jumps are
sufficiently long, or a mixture of both.

Finally, we end with an interesting observation that does not appear
obvious.  One might, instead of Eq.~(\ref{theintegralequation}), have been
motivated to write the integral equation
\begin{widetext}
\begin{equation}
w(\mathbf{r},t|\mathbf{r}_0,0)=\int d\mathbf{r'}\,
\int_0^{t} dt'\,\Psi(\mathbf{r'}-\mathbf{r}_0,t')
\Xi(\mathbf{r'},\mathbf{r}_0; t',0)
w(\mathbf{r},t|\mathbf{r'},t')
+\Phi(t)\Xi(\mathbf{r}_0,\mathbf{r}_0;t,0)\delta(\mathbf{r}
-\mathbf{r}_0).
\label{theintegralequationalt}
\end{equation}
\end{widetext}
Again in the language of Hughes~\cite{Hughes1}, this corresponds to a
``partition over the first step.'' What is interesting is that we are not able
to arrive at any reasonable fractional diffusion equation starting from this
integral equation, in any case not by the methods followed in this paper for
Eq.~(\ref{theintegralequation}) even though one might have expected a certain
symmetry to the situation. This, too, is a question to be explored further.

\acknowledgments     
The authors thank Rafael Borrego for his careful reading of the manuscript and
his resultant suggestions for improvement. This
work was partially supported by the Ministerio de Ciencia y
Tecnolog\'{\i}a (Spain) through Grant No. FIS2007-60977, by the Junta de
Extremadura (Spain)
through Grant No. GRU09038, and by the
National Science Foundation under grant No. PHY-0855471.

\end{document}